\documentclass[a4paper, 10 pt, conference]{hsmr} 
\usepackage[utf8]{inputenc}
\IEEEoverridecommandlockouts                       
\usepackage{mathtools}
\usepackage{array}
\usepackage{geometry}
\usepackage[labelsep=space]{caption}
\usepackage{newtxmath,newtxtext}
\usepackage{authblk}
\usepackage{pgfplots}
\usepackage{graphicx}
\usepackage[colorlinks,urlcolor=black]{hyperref} 
\usepackage{newtxtext,newtxmath}
\usepackage{xcolor}
\usepackage{tikz}
\usetikzlibrary{math,shapes,calc,decorations.text}

\usepackage[backend=biber,
            hyperref=true,
            url=false,
            isbn=false,
            doi=false,
            backref=false,
            style=ieee,
            natbib=true,
            mincitenames=1,
            maxcitenames=1,
            citestyle=numeric-verb,
            sorting=none,
            block=none]{biblatex}
\bibliography{./main.bib}

\pgfplotsset{compat=newest}

\title{\LARGE \bf
A Modular Edge Device Network for Surgery Digitalization
}

\author{Vincent Schorp$^1$}
\author{Frédéric Giraud$^{1}$}
\author{Gianluca Pargätzi$^{2}$}
\author{Michael Wäspe$^{2}$}
\author{Lorenzo von Ritter-Zahony$^{2}$}
\author{\\Marcel Wegmann$^{2}$}
\author{Nicola A. Cavalcanti$^{1}$}
\author{John Garcia Henao$^{1}$}
\author{Nicholas Bünger$^{1}$}
\author{\\Dominique Cachin$^{2}$}
\author{Sebastiano Caprara$^{1}$}
\author{Philipp Fürnstahl$^{1}$}
\author{Fabio Carrillo$^{1}$}

\affil{\textit{$^{1}$Balgrist University Hospital, University of Zurich, Zurich, Switzerland}\\ \textit{$^{2}$Institute of Embedded Systems, ZHAW School of Engineering, Winterthur, Switzerland}\\ \textit{fabio.carrillo@balgrist.ch}}

\begin{document}

\maketitle
\thispagestyle{empty}
\pagestyle{empty}

\section*{INTRODUCTION}
\label{sec:intro}

Future surgical care will increasingly rely on collaboration between caregivers, patients, technology, and information systems, driven by data science. The operating room (OR) of the future will be fully interconnected, providing surgical teams with real-time information for improved decision-making and efficiency. However, integration of machine learning in interventional medicine remains slow due to limited digitization and standardization of patient data \cite{maier-hein_surgical_2017}.

Automated multimodal data acquisition and processing can address these challenges by enabling surgical digital twins, foundation models, and advanced clinical diagnostics, thereby augmenting surgical capabilities. Additionally, such data will empower medical robots and enable remote participation, learning, and eventually teleoperation. However, real-time synchronized data capture remains problematic due to diverse device interfaces, protocols, and extensive cabling, compromising OR sterility and ergonomics. To our knowledge, no commercially available solution currently provides a comprehensive pipeline — covering hardware, middleware, and interfaces — for real-time intraoperative data acquisition.


Recent projects at OR-X \cite{orx_min}, a translational hub for surgical research, illustrate these challenges. For instance, a surgical digitization system by \textcite{Hein_2024_CVPR_min} involved multiple interconnected cameras, resulting in a complex setup. Similarly, a robotic ultrasound (US) scanning method proposed by \textcite{Cavalcanti_2024_min} highlighted the necessity to integrate robots, US scanners, cameras, and tracking devices for adaptive scanning and real-time anatomical reconstruction.

\begin{figure}[h]
\centering
\includegraphics[width=\columnwidth]{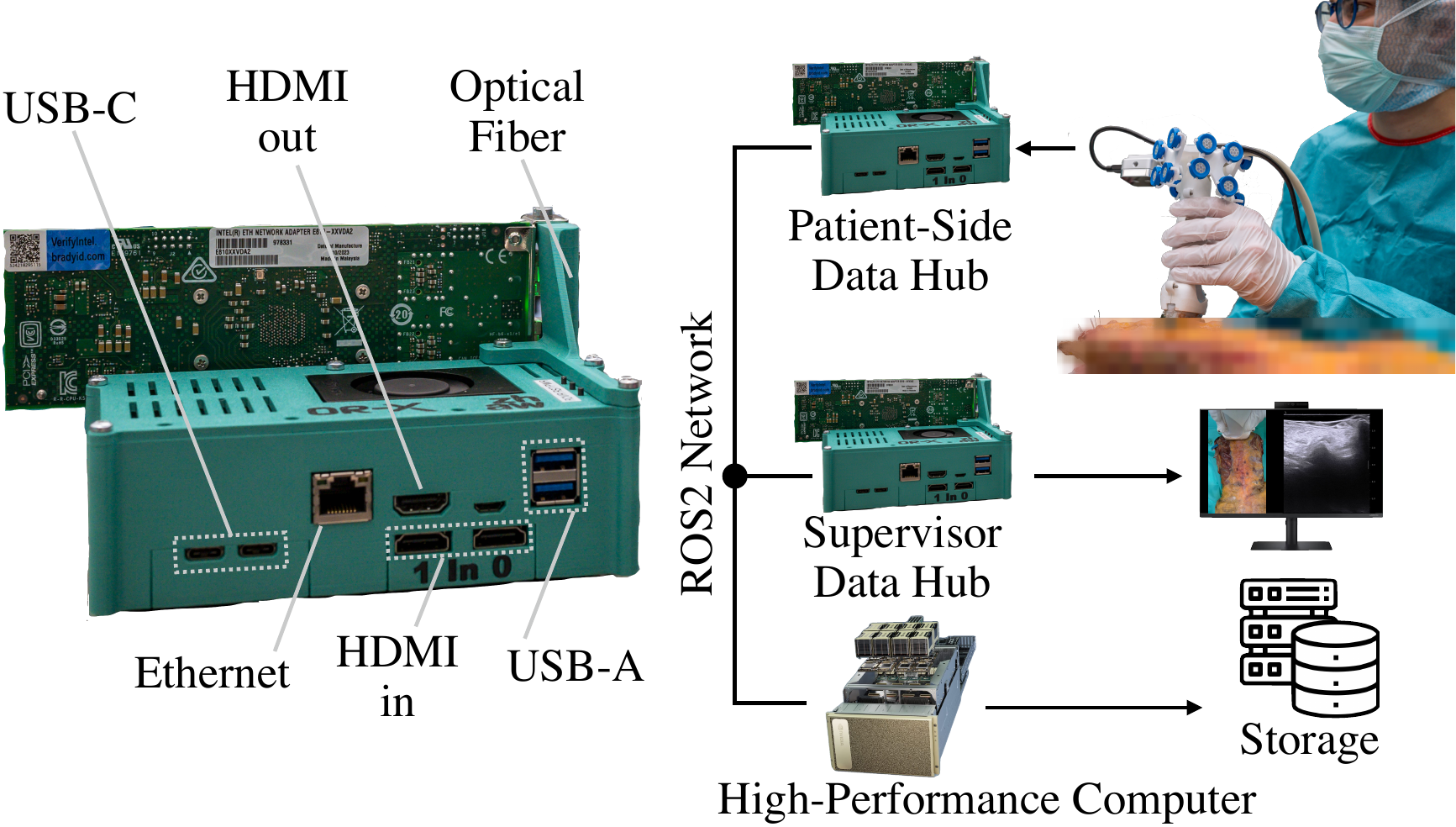}
\caption{\textbf{Data Hub Network.} \emph{Left:} Data Hub (DH) with all ports. \emph{Right:} Experimental setup comprised of a patient-side DH connected to a US scanner, a pose tracking device, and an RGB-D camera; a supervisor DH to configure and monitor the experiment; and a high-performance computer for data storage and processing. Further DH can be added to the OR-X Network. The real-time transfer of the synchronized data is achieved using a ROS2 network.}
\vspace{-0.25in}
\label{fig_complete}
\end{figure}


We propose a high-performance computer network comprising strategically placed edge devices (Data Hubs, DH) interconnected via optical fiber through a network switch (Fig. \ref{fig_complete}). These DHs facilitate seamless integration and synchronized multimodal data capture from medical devices, sensors, and robotic systems, supporting real-time processing for machine learning applications. The system leverages an NVIDIA DGX computer for state-of-the-art data processing and offers centralized monitoring and control via an intuitive user interface.

We validated this edge device network in an ongoing US-based 3D anatomical reconstruction project, representing a challenging, realistic surgical scenario involving medical imaging, pose tracking, and RGB-D imaging. Our system demonstrated the capability to simultaneously collect and store high-resolution, high-frame-rate data from multiple devices, while providing an intuitive interface for configuring and monitoring experiments.

The contributions of this work include:
\begin{itemize}
    \item A modular high-performance computer network enabling high-bandwidth data acquisition and real-time processing.
    \item An edge device capable of interfacing with multimodal sensors and robotic systems for supporting surgical research. 
    \item An intuitive software solution for managing a network of sensors and devices.
    \item Experimental validation in a realistic surgical environment, showcasing the system’s utility in complex procedures.
\end{itemize}

\section*{MATERIALS AND METHODS}

\subsection*{Data Hub Hardware}
The DH is an edge computing device tailored for surgical environments with multimodal bidirectional ports and embedded computing capabilities. It is based on an NVIDIA Jetson Orin NX featuring an NVIDIA Ampere GPU and augmented with an Intel E810 network adapter to enable high-throughput optical fiber communication. The hardware includes two HDMI input ports, one HDMI output port, one Ethernet port, two optical fiber interfaces, two USB-C 3.2 ports, and two USB-A 3.2 ports, as shown in Fig \ref{fig_complete}. While a network of DHs can operate independently, an NVIDIA DGX computer equipped with eight A100 GPUs was added to enable state-of-the-art data processing and large-scale storage. 

\subsection*{Data Hub Software}
The software architecture of the DH is built on the Isaac ROS framework \cite{isaac_ros}, which standardizes communication with various devices. Each supported device is encapsulated within a Docker image containing the necessary drivers, SDKs, and a ROS 2 \cite{ros2} package. 
Currently, the DH supports multiple devices including ZED Mini and Intel RealSense cameras (via USB-C/USB-A); Atracsys FusionTrack (via Ethernet); KUKA LBR Medical Robot (via Ethernet); and Aixplorer Ultrasound systems (via HDMI input). To support high-throughput data transfer, GPU-enabled H.264 image compression and decompression modules have been integrated into the system. To facilitate the deployment of complex experimental setups, one DH is designated as the central control unit. An adaptable user interface was created as a single entrypoint for convenient device configuration and real-time control and monitoring. We programmed the interface using Streamlit \cite{streamlit_min} and Docker.


\subsection*{Experimental Setup}
\label{sec:exp_setup}
The DH network is evaluated on an US scanning experiment, depicted in Fig~\ref{fig_complete}, on an ex-vivo human spine anatomy\footnote{BASEC N° 2021-01196 approved by the Ethical Committee of the Kanton of Zurich}. The data collection leverages a SuperSonic Aixplorer Ultimate handheld US scanner, an Atracsys FusionTrack 500 visual tracking device, and an Intel Realsense D405 RGB-D camera. The procedure aims to collect the raw data required to reconstruct the anatomy from the US images and a mesh of the visible surface from the RGB-D data. A network of two DHs was utilized in this procedure. The patient-side DH was used to connect the three sensors mentioned above, which feature HDMI, Ethernet, and USB-C interfaces, respectively. The monitoring DH was utilized to configure, start, stop, and monitor the data recording procedure. During dataset acquisition, the data is stored in real-time on the NVIDIA DGX storage space, where it is post-processed later.

\section*{RESULTS}
The proposed network of edge devices allows the recording of the data from all three sensors at their maximum frame rate and resolution. The results can be seen in Table~\ref{tab:exp_device_table}. The scan of the spine took a total of 78 seconds, during which 4622 US images, 2283 images each for RGB and depth, and 12434 3D poses were acquired, amounting to a total of 6.2~GB.

\begin{table}[h!]
\vspace{-0.0in}
\centering
\caption{Frame rates and resolutions achieved during the experiment.}

\begin{tabular}{|l|l|l|}
\hline
\textbf{Device} & \textbf{Connector} & \textbf{Data Format and FPS} \\ \hline
Aixplorer US scanner & HDMI & RGB 1080p @ 60.2 FPS \\ \hline
Atracsys Fusion 500 & Ethernet & 3D Pose @ 200.8 FPS \\ \hline
Intel Realsense D405 & USB-C & RGB 720p @ 29.6 FPS \\ \hline
Intel Realsense D405 & USB-C & Depth 720p @ 29.6 FPS \\ \hline
\end{tabular}
\vspace{-0.1in}
\label{tab:exp_device_table}
\end{table}
\section*{DISCUSSION}


This proof of concept demonstrates that the proposed network reliably acquires synchronized data streams from multiple devices, at the highest possible resolution and frame rate. DHs enable an efficient, clutter-free setup. The user interface allows straightforward configuration and monitoring of the experiment, while all data is stored on the DGX server for offline analysis and machine learning training.

We validated the system with three common sensing devices, representing a realistic intraoperative use case. Future work will expand device compatibility to support a broader range of setups, including robotic systems requiring real-time, closed-loop control.  

This platform lays the groundwork for scaling to larger, more complex configurations. Expanding the number of DHs and connected devices will necessitate increased data throughput and improved real-time stream monitoring. The current OR-X infrastructure has been dimensioned to support these technical requirements, but its evaluation is subject to future work.  

Ultimately, this system forms the technical backbone of our broader surgical digitization initiative. Future deployments may distribute DHs across the OR as distributed connected devices, eventually enabling efficient intraoperative data collection for machine learning training, real-time inference, remote participation, and teleoperation. 




\printbibliography

@InProceedings{Hein_2024_CVPR_min,
    author    = {Hein, Jonas and Giraud, Fr\'ed\'eric and Calvet, Lilian and Schwarz, Alexander and Cavalcanti, Nicola Alessandro and Prokudin, Sergey and Farshad, Mazda and Tang, Siyu and Pollefeys, Marc and Carrillo, Fabio and F\"urnstahl, Philipp},
    title     = {Creating a Digital Twin of Spinal Surgery: A Proof of Concept},
    booktitle = {CVPR},
    year      = {2024},
}

@inproceedings{Cavalcanti_2024_min,
 author = {
Cavalcanti, Nicola A. and Carrillo, Fabio and Li, Ruixuan and Van Assche, Kaat and Davoodi, Ayoob and Tummers, Matthias and Huber, Martin and Teyssere,
François and Pérez Velásquez, Jorge Andrés and Massalimova, Aidana and Laux, Christoph J and Sutter, Reto and Farshad, Mazda and Borghesan,
Gianni and Denis, Katleen. and Morel, Guillaume and Chandanson, Thibault and Vercauteren, Tom and Vander Poorten, Emmanuel and Fürnstahl, Philipp},
 booktitle = {CLINICCAI},
 doi = {10.5281/zenodo.13134982},
 publisher = {Zenodo},
 title = {Bridging innovation and practice: the journey of FAROS from technical design to in-vivo animal validation},
 year = {2024}
}

@misc{orx_min,
  howpublished = {\url{https://or-x.ch/en/translational-center-for-surgery/}},
}

@misc{streamlit_min,
  howpublished = {\url{https://streamlit.io}},
}

@misc{isaac_ros,
  howpublished = {\url{https://developer.nvidia.com/isaac/ros}},
}

@misc{ros2,
  howpublished = {\url{https://docs.ros.org/en/humble/index.html}},
}

@article{maier-hein_surgical_2017,
	title = {Surgical data science for next-generation interventions},
	volume = {1},
	issn = {2157-846X},
	url = {https://doi.org/10.1038/s41551-017-0132-7},
	doi = {10.1038/s41551-017-0132-7},
	number = {9},
	journal = {Nature Biomedical Engineering},
	author = {Maier-Hein, Lena and Vedula, Swaroop S. and Speidel, Stefanie and Navab, Nassir and Kikinis, Ron and Park, Adrian and Eisenmann, Matthias and Feussner, Hubertus and Forestier, Germain and Giannarou, Stamatia and Hashizume, Makoto and Katic, Darko and Kenngott, Hannes and Kranzfelder, Michael and Malpani, Anand and März, Keno and Neumuth, Thomas and Padoy, Nicolas and Pugh, Carla and Schoch, Nicolai and Stoyanov, Danail and Taylor, Russell and Wagner, Martin and Hager, Gregory D. and Jannin, Pierre},
	month = sep,
	year = {2017},
	pages = {691--696},
}


\end{document}